\documentclass[showpacs,twocolumn,typeset,superscriptaddress]{revtex4}
\usepackage{amsfonts}
\usepackage{amsmath}
\usepackage{amssymb}
\usepackage{graphicx}
\usepackage{dcolumn}
\usepackage{bm}

\begin{document}
\preprint{APS}

\title{Entanglement-swapping for $X$-states demands threshold values}
\author{Ariana Mu\~noz}
\affiliation{Departamento de F\'{\i}sica, Universidad de Concepci\'{o}n, Barrio Universitario, Casilla 160-C, Concepci\'{o}n, Chile.}
\author{Gesa Gr\"uning}
\affiliation{Departamento de F\'{\i}sica, Universidad de Concepci\'{o}n, Barrio Universitario, Casilla 160-C, Concepci\'{o}n, Chile.}
\affiliation{Ruprecht Karls Universit\"at Heidelberg, Heidelberg, Germany.}
\author{Luis Roa}\email{lroa@udec.cl}
\affiliation{Departamento de F\'{\i}sica, Universidad de Concepci\'{o}n, Barrio Universitario, Casilla 160-C, Concepci\'{o}n, Chile.}

\date{\today}

\begin{abstract}
The basic entanglement-swapping scheme can be seen as a process which allows to redistribute the Bell states' properties between different pairs of a four qubits system.
Achieving the task requires performing a von Neumann measurement, which projects a pair of factorized qubits randomly onto one of the four Bell states.
In this work we propose a similar scheme, by performing the same Bell-von Neumann measurement over two local qubits, each one initially being correlated through an $X$-state with a spatially distant qubit.
This process swaps the $X$-feature without conditions, whereas the input entanglement is partially distributed in the four possible outcome states under certain conditions.
Specifically, we obtain two threshold values for the input entanglement in order for the outcome states to be nonseparable.
Besides, we find that there are two possible amounts of outcome entanglement, one is greater and the other less than the input entanglement;
the probability of obtaining the greatest outcome entanglement is smaller than the probability of achieving the least one.
In addition, we illustrate the distribution of the entanglement for some particular and interesting initial $X$-states.
\end{abstract}

 \maketitle

\section{Introduction}

Quantum correlations are in the heart of a number of quantum effects such as quantum nonlocality \cite{AE,books,JFC,MBP,RH,CHB,GBR}, the speedup \cite{EK,AD}, and unambiguous state discrimination \cite{AP,LR}.
Even more, quantum correlations are the key ingredient for implementing processes between distant particles, for instance, the teleportation of unknown state \cite{CHB,CHBKB} and the remote state preparation \cite{CHBDPDV}.
Thus, quantum correlations become fundamental resources for attaining novel quantum effects.
In this context, quantum correlations have been well characterized hitherto by the so-called entanglement and quantum discord \cite{WKW,LH}.
The discord can be generated by starting from a separable classical state and using local operations and classical communication \cite{JFC,MBP};
whereas the preparation of a bipartite entangled state requires a direct interaction of the particles at their common past.
The entanglement-swapping \cite{MZ} procedure does not infringe the above statement since it simply redistributes quantum correlation between different parts of two composed systems.
The advantage of this process lies in the fact that two factorized parts can acquire entanglement even though they are away from each other and from the other parts.
Numerous are the experimental realizations reported in the literature which account for some application of the entanglement-swapping process \cite{MR,CB};
recently it has been demonstrated that entanglement can be swapped to timelike separated quantum systems \cite{EM}.

In this work we propose a generalization of the basic entanglement-swapping scheme by considering two pairs of qubits being initially correlated via $X$-states instead of pure ones.
We start by showing that the basic entanglement-swapping process redistributes the entanglement when we consider pure and partially entangled states.
In this case, two of the four outcome states are maximally entangled and the other two have entanglement smaller than the input one.
This motivates us to study the process by considering two pairs of qubits prepared initially in mixed states,
specifically the $X$-types for which there is an analytical expression for the concurrence \cite{WKW,TY}.
We focus with special interest on analyzing the entanglement redistribution accomplished in this process.

Further and major motivations for studying $X$-states stem from the fact that they are usually encountered in different areas;
for instance, some decoherence mechanisms can take two qubits to an $X$-state \cite{Yu04,Marcelo06,Qi-liang},
two two-level atoms in the Tavis-Cummings model can reach a dynamics $X$-state \cite{JCRLRCL},
the states with minimum and the states with maximum discord for a fixed entanglement value are $X$-states \cite{AA},
in condensed matter the ground state of two $Z_2$-symmetry sites of a lattice is represented by an $X$-state as well \cite{condenceM}.

This article is organized as follows: in Sec. \ref{pure state} we expose the principal aspects of the entanglement-swapping process with pure states.
In Sec. \ref{x-state} we describe the main properties of an $X$-state.
In Sec. \ref{2xstate} we address the entanglement-swapping scheme starting with two different $X$-states.
In Sec. \ref{x state entanglement} we analyze the entanglement swapping in the special case where the two initial $X$-states are equal.
Some interesting and particular examples are studied in Sec. \ref{rhodifferent} where the entanglement is partially redistributed.
Finally, in the last section we summarize our results.

\section{The basic entanglement-swapping scheme} \label{pure state}

\begin{figure}[t]
\includegraphics[angle=360,width=0.40\textwidth]{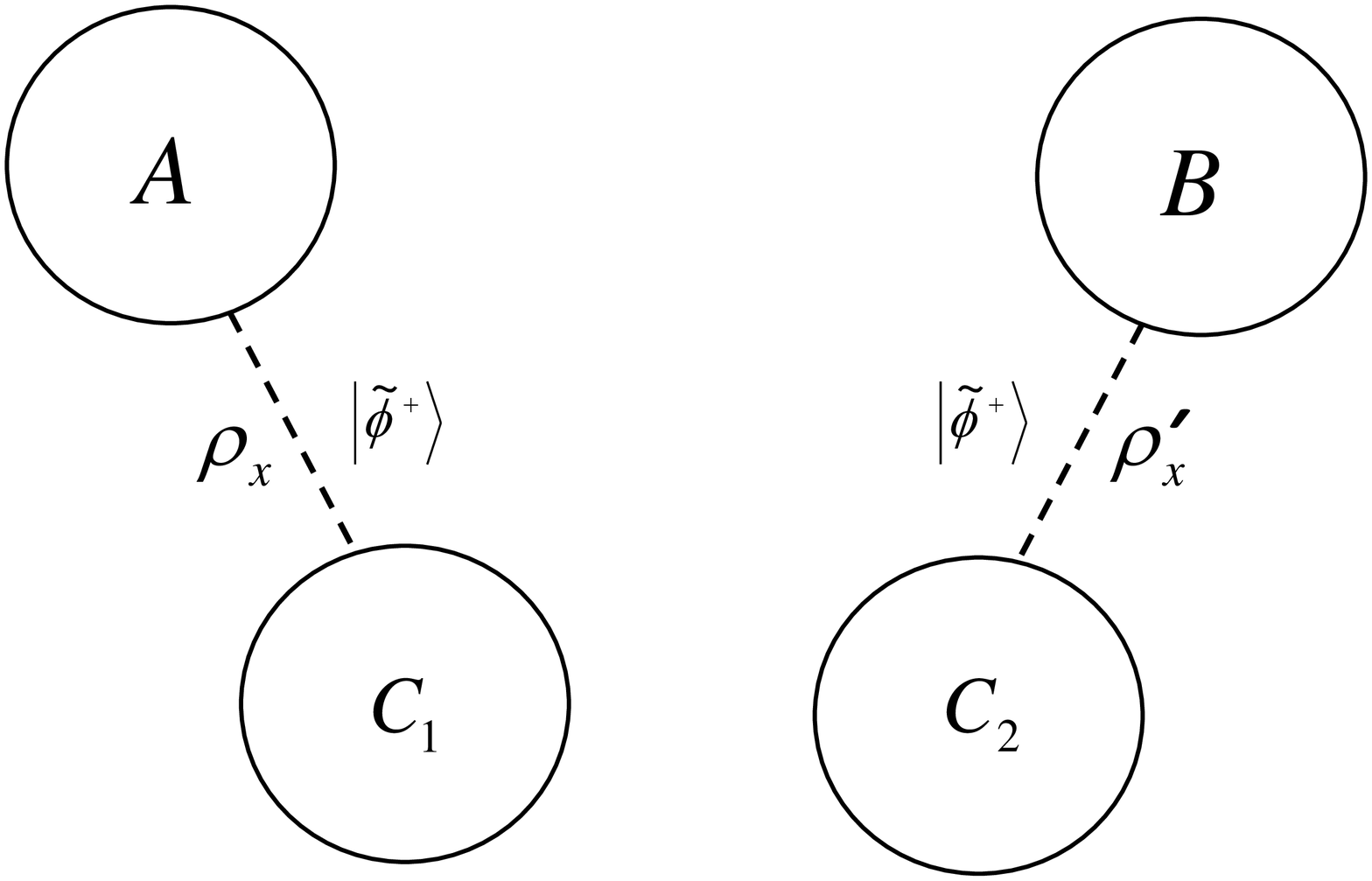}
\caption{Entanglement-swapping scheme.
The qubits $C1$ and $C2$ are in the same laboratory, whereas $A$ and $B$ are spatially remote from each other and from qubits $C1$ and $C2$.}
\label{figure0}
\end{figure}

In this section we describe the principal aspects of the entanglement-swapping schemes with pure states.
Let us consider the simplest case of two pairs of qubits $A,C1$ and $B,C2$ prepared in the Bell states $|\phi^+_{A,C1}\rangle$ and $|\phi^+_{B,C2}\rangle$; see Fig. \ref{figure0}.
The entanglement-swapping protocol can readily be read out of the identity
\begin{eqnarray}
\!\!\!|\phi^+_{A,C1}\rangle|\phi^+_{B,C2}\rangle\!\! &=&\!\!\!\frac{1}{2}\left(|\phi^+_{A,B}\rangle|\phi^+_{C1,C2}\rangle+|\phi^-_{A,B}\rangle|\phi^-_{C1,C2}\rangle \right.\nonumber\\
\!\!\!&&\!\!+\left.\!|\psi^+_{A,B}\rangle\!|\psi^+_{C1,C2}\rangle\!+\!|\psi^-_{A,B}\rangle\!|\psi^-_{C1,C2}\rangle\!\!\right)\!, \label{swapingpure}
\end{eqnarray}
where $|\phi^{\pm}_{U,V}\rangle=(|0_U\rangle|0_V\rangle\pm|1_U\rangle|1_V\rangle)/\sqrt{2}$ and $|\psi^{\pm}_{U,V}\rangle=(|0_U\rangle|1_V\rangle\pm|1_U\rangle|0_V\rangle)/\sqrt{2}$
are the Bell states of the systems $U$ and $V$, and $\{|0\rangle,|1\rangle\}$ are the eigenstates of the Pauli operator $\sigma_z$ of the respective system.
From (\ref{swapingpure}) we realize that by implementing a measurement process, which projects the qubits $C1$ and $C2$ onto one of their Bell states,
then the pair of qubits $A,B$ are projected also onto one of their Bell states.
Each one of the four results has the same probability $1/4$.
Therefore, by projecting onto the Bell basis of the qubits $C1,C2$ the entanglement contained in the two pairs $A,C1$ and $B,C2$ is redistributed to the pair $A,B$,
even though there is no interaction between them.

\begin{figure}[t]
\includegraphics[angle=360,width=0.40\textwidth]{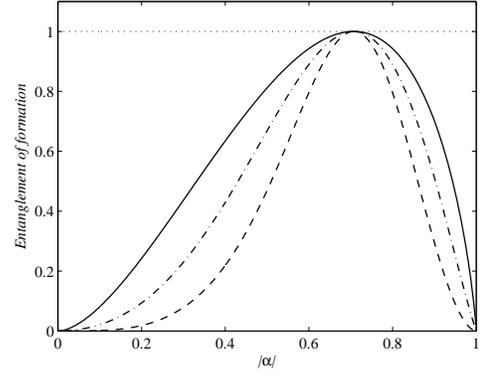}
\caption{Entanglement of formation as functions of $|a|$ of the initial states $|\tilde{\phi}^+_{U,V}\rangle$ (solid line)
and of the outcome ones $|\ddot{\phi}^\pm_{A,B}\rangle$ (dashed line) and $|\psi^\pm_{A,B}\rangle$ (dotted line).
The dot-dashed line corresponds to the average entanglement $\bar{E}$ of the four possible outcome states $|\ddot{\phi}^\pm_{A,B}\rangle$ and $|\psi^\pm_{A,B}\rangle$.}
\label{figure1}
\end{figure}

When the two pairs of qubits $A,C1$ and $B,C2$ initially are in partially entangled pure states the identity (\ref{swapingpure}) becomes
\begin{eqnarray}
\!\!\!|\tilde{\phi}^+_{A,C1}\rangle|\tilde{\phi}^+_{B,C2}\rangle\!\!\! &=&\!\!\!\!\sqrt{p_\phi}\!\left(|\ddot{\phi}^+_{A,B}\rangle\!|\phi^+_{C1,C2}\rangle\!+\!|\ddot{\phi}^-_{A,B}\rangle\!|\phi^-_{C1,C2}\rangle\!\right) \nonumber\\
\!\!\!\!\!&&\!\!\!\!+ab\!\left(\!\!|\psi^+_{A,B}\rangle\!|\psi^+_{C1,C2}\rangle\!+\!|\psi^-_{A,B}\rangle\!|\psi^-_{C1,C2}\rangle\!\!\right)\!\!, \label{swapingpure2}
\end{eqnarray}
where
\begin{eqnarray*}
|\tilde{\phi}^+_{U,V}\rangle&=&a|0_U\rangle|0_V\rangle+b|1_U\rangle|1_V\rangle, \\
|\ddot{\phi}^\pm_{A,B}\rangle&=&\frac{a^2|0_A\rangle|0_B\rangle\pm b^2|1_A\rangle|1_B\rangle}{\sqrt{|a|^4+|b|^4}},\\
p_\phi &=&\frac{|a|^4+|b|^4}{2},
\end{eqnarray*}
with $|a|^2+|b|^2=1$.
Note that by projecting the qubits $C1$ and $C2$ onto one of the Bell states $|\psi^\pm_{C1,C2}\rangle$ the pair $A,B$
is projected as well onto one of the Bell states $|\psi^\pm_{A,B}\rangle$ with probability $p_\psi=|a|^2|b|^2$, otherwise
when the pair $C1,C2$ is projected onto one of the Bell states $|\phi^\pm_{C1,C2}\rangle$ the composed system $A,B$ also
is projected onto one of the partially entangled states $|\ddot{\phi}^\pm_{A,B}\rangle$ with probability $p_\phi\geq p_\psi$.
Thus, even though the initial entanglement $E(|\tilde{\phi}^+_{U,V}\rangle)$ is not maximal,
there are two possible outcome states maximally entangled.
However, the other two resulting states have amount of entanglement $E(|\ddot{\phi}^\pm_{A,B}\rangle)$ smaller than the initial;
even more, the average outcome entanglement $\bar{E}$ is smaller than the initial one $E(|\tilde{\phi}^+_{U,V}\rangle)$.
We illustrate that behavior of the redistributed entanglement in Fig. \ref{figure1}, as functions of $|a|$.
We note that the four outcome states are maximally entangled only for $|a|=1/\sqrt{2}$, which is the case of the identity (\ref{swapingpure}).
It is worth emphasizing three affairs;
i) the outcome entanglement is maximal with probability $2|a|^2|b|^2$, which vanishes only at the end values $|a|=0$, $1$ where there is no input entanglement,
ii) the probability of increasing the entanglement is always smaller than the one for decreasing it, since $2|a|^2|b|^2\leq |a|^4+|b|^4$,
iii) for having entangled outcome states it is required to have initial entanglement different from zero only.
Finally, we want to emphasize that there is an asymmetry in the distribution of the entanglement over the four outcome states, which in the identity (\ref{swapingpure2}) 	
looks with favor on the $|\psi^\pm_{A,B}\rangle$ outcome Bell states.

\section{$X$-state main features} \label{x-state}

Here we review succinctly the properties of a general $X$-state $\hat{\rho}_{a,b}$ of two qubits $a$ and $b$.
In the logic basis $\{|0_a\rangle|0_b\rangle,|0_a\rangle|1_b\rangle,|1_a\rangle|0_b\rangle,|1_a\rangle|1_b\rangle\}$ the state is represented by the matrix
\begin{equation}
\hat{\rho}_{a,b}\equiv\left(
\begin{array}{cccc}
\rho_{11}  & 0         & 0         &\rho_{14}\\
0          & \rho_{22} & \rho_{23}& 0 \\
0          & \rho_{32}& \rho_{33} & 0 \\
\rho_{41}& 0         & 0         & \rho_{44}
\end{array}
\right). \label{Xstate}
\end{equation}
In order for (\ref{Xstate}) to describe a quantum state, the matrix elements must satisfy the normalization $\sum_{j=1}^{4}\rho_{jj}=1$ and the positivity conditions,
\begin{equation}
|\rho_{14}|\leq\sqrt{\rho_{11}\rho_{44}}\quad\text{and}\quad |\rho_{23}|\leq\sqrt{\rho_{22}\rho_{33}}. \label{positivity}
\end{equation}
In what follows we consider \emph{fixed diagonal elements} and we do the analysis in terms of the off-diagonal elements.
The off-diagonal elements account for the coherence degree inside two orthogonal subspaces, say,
$\mathcal{H}_{00,11}$ spanned by the basis $\{|0_a\rangle|0_b\rangle,|1_a\rangle|1_b\rangle\}$ and $\mathcal{H}_{01,10}$ spanned by $\{|0_a\rangle|1_b\rangle,|1_a\rangle|0_b\rangle\}$.
For instance, when $\rho_{14}=0$ there is absolute decoherence into $\mathcal{H}_{00,11}$, while for $|\rho_{14}|=\sqrt{\rho_{11}\rho_{44}}$ there is a pure state into $\mathcal{H}_{00,11}$.
The $\rho_{23}$ element has similar meaning in the subspace $\mathcal{H}_{01,10}$.
Accordingly, depending on the moduli $|\rho_{14}|$ and $|\rho_{23}|$ the $X$-state (\ref{Xstate}) goes from an incoherent superposition of the four factorized logic states
to an incoherent superposition of two partially entangled pure states.

There are two special features of an $X$-state:
i) the $X$ form in itself which arises because it populates the two subspaces $\mathcal{H}_{00,11}$ and $\mathcal{H}_{01,10}$
without having off-diagonal terms between the elements of those subspaces,
ii) the entanglement between the two involved systems.

The entanglement can be evaluated by the concurrence \cite{WKW}, which for the (\ref{Xstate}) state is given by \cite{TY},
\begin{equation}
\mathcal{C}_{in}=2\max\{0,|\rho_{14}|-\sqrt{\rho_{22}\rho_{33}},|\rho_{23}|-\sqrt{\rho_{11}\rho_{44}}\}.
\end{equation}
Thus, there is nonzero entanglement when one of the two following inequalities is satisfied:
\begin{equation}
|\rho_{14}|>\sqrt{\rho_{22}\rho_{33}}, \qquad|\rho_{23}|>\sqrt{\rho_{11}\rho_{44}},\label{rhooverrho}
\end{equation}
otherwise entanglement is absent.
Complementing inequalities (\ref{rhooverrho}) with those for positivity (\ref{positivity}),
leads to the following two inequations for having entanglement:
\begin{subequations}\begin{eqnarray}
|\rho_{23}|\leq\sqrt{\rho_{22}\rho_{33}}&<&|\rho_{14}|\leq\sqrt{\rho_{11}\rho_{44}}, \label{eqFORe1}\\
|\rho_{14}|\leq\sqrt{\rho_{11}\rho_{44}}&<&|\rho_{23}|\leq\sqrt{\rho_{22}\rho_{33}}, \label{eqFORe2}
\end{eqnarray}\label{eqFORe}\end{subequations}
clearly only one or none of them can be fulfilled.
We extract in the following three sentences what the inequalities (\ref{eqFORe}) are saying to us:
\begin{itemize}
  \item[i)] For $\mathcal{C}_{in}\neq0$ it is necessary that $\rho_{11}\rho_{44}\neq\rho_{22}\rho_{33}$.
  \item[ii)] For $\mathcal{C}_{in}\neq0$ it is necessary and sufficient to have a certain nonzero coherence degree inside only one subspace $\mathcal{H}_{00,11}$ or $\mathcal{H}_{01,10}$, i.e., $|\rho_{ij}|>\sqrt{\rho_{nn}\rho_{mm}}$, with $i\neq j\neq n \neq m$.
  \item[iii)] The coherence degree into the other subspaces, $|\rho_{nm}|$, does not contribute to $\mathcal{C}_{in}$; even more,
  $|\rho_{nm}|$ can be zero and even then it does not affect the amount of entanglement of the $X$-state.
  \item[iv)] The entanglement dies just when the $\max\{|\rho_{14}|,|\rho_{23}|\}=\min\{\sqrt{\rho_{11}\rho_{44}},\sqrt{\rho_{22}\rho_{33}}\}$.
\end{itemize}
We stress that, for fixed diagonal elements, the maximum amount of entanglement is reached by satisfying condition i) and having total coherence into only one subspace
$\mathcal{H}_{00,11}$ or $\mathcal{H}_{01,10}$ in agreement with condition ii).
The inequations (\ref{eqFORe}) define, in the composed Hilbert space, the border between the entangled and the separable $X$-states.

\section{$X$-states swapping} \label{2xstate}

Let us consider two pairs of qubits $A$,$C1$ and $B$,$C2$.
The qubits $C1$ and $C2$ are in the same laboratory, in this way local and joint operations can be applied on them.
The $A$ and $B$ qubits are spatially remote from each other and from qubits $C1$ and $C2$, accordingly, it is not possible to perform joint operations onto $A$ and $B$; see Fig. \ref{figure0}.
We assume that, at the common past, both pairs of qubits $A,C1$ and $B,C2$ were prepared in the $X$-states $\hat{\rho}_{A,C1}$ and $\hat{\rho}_{B,C2}$, which we call input states.
In the respective logic bases $\{|0_A\rangle|0_{C1}\rangle,|0_A\rangle|1_{C1}\rangle,|1_A\rangle|0_{C1}\rangle,|1_A\rangle|1_{C1}\rangle\}$ and $\{|0_B\rangle|0_{C2}\rangle,|0_B\rangle|1_{C2}\rangle,|1_B\rangle|0_{C2}\rangle,|1_B\rangle|1_{C2}\rangle\}$ the input states are represented by the matrices
\begin{equation}
\begin{small}
\hat{\rho}_{A,C1}\!\equiv\!\!\left(\!\!
\begin{array}{cccc}
\rho_{11}\!\! & 0 & 0 & \!\!\rho_{14} \\
0 &\!\! \rho_{22} & \rho_{23}\!\! & 0 \\
0 &\!\! \rho_{32} & \rho_{33}\!\! & 0 \\
\rho_{41}\!\! & 0 & 0 & \!\!\rho_{44}
\end{array}
\!\!\right)\!\!,\hspace{0.06in}
\hat{\rho}_{B,C2}\!\equiv\!\!\left(\!\!
\begin{array}{cccc}
\rho_{11}^{\prime } \!\!& 0 & 0 & \!\!\rho_{14}^{\prime } \\
0 & \!\!\rho_{22}^{\prime } & \rho_{23}^{\prime }\!\! & 0 \\
0 &\! \!\rho_{32}^{\prime } & \rho_{33}^{\prime }\!\! & 0 \\
\rho_{41}^{\prime }\!\! & 0 & 0 & \!\!\rho_{44}^{\prime }
\end{array}
\!\!\right)\!\!. \!\!\label{Xstates}
\end{small}
\end{equation}
Now we consider that a projective measurement is performed over the qubits $C1$ and $C2$.
Specifically, onto the factorized state $\hat{\rho}_{A,C1}\otimes\hat{\rho}_{B,C2}$ is applied a von Neumann measurement, which projects the pair $C1,C2$
onto one of the four Bell states
\begin{subequations}
\begin{eqnarray}
|\phi^\pm_{C1,C2}\rangle&=&\frac{1}{\sqrt{2}}\left(|0_{C1}\rangle|0_{C2}\rangle\pm |1_{C1}\rangle|1_{C2}\rangle\right), \label{Bell1}\\
|\psi^\pm_{C1,C2}\rangle&=&\frac{1}{\sqrt{2}}\left(|0_{C1}\rangle|1_{C2}\rangle\pm |1_{C1}\rangle|0_{C2}\rangle\right). \label{Bell2}
\end{eqnarray}\label{Belltipe}\end{subequations}%
Consequently the pair $A,B$ also experiments a projection onto one of the states $\hat{\rho}^{\phi^\pm}_{AB}$ or $\hat{\rho}^{\psi^\pm}_{AB}$, which we call outcome states.
In the logic basis $\{|0_A\rangle|0_B\rangle,|0_A\rangle|1_B\rangle,|1_A\rangle|0_B\rangle,|1_A\rangle|1_B\rangle\}$ the outcome states are represented by the matrices
\begin{widetext}
\begin{subequations}
\begin{equation}
\hat{\rho}_{AB}^{\phi^{\pm }} \equiv \frac{1}{N_{\phi }}\left(
\begin{array}{cccc}
\rho _{11}\rho _{11}^{\prime }+\rho _{22}\rho _{22}^{\prime } & 0 & 0 & \pm \left(\rho_{14}\rho _{14}^{\prime }+ \rho _{23}\rho _{23}^{\prime }\right)  \\
0 & \rho _{11}\rho _{33}^{\prime }+\rho _{22}\rho _{44}^{\prime } & \pm \left(\rho_{14}\rho _{32}^{\prime }+\rho _{23}\rho _{41}^{\prime }\right)  & 0 \\
0 & \pm \left( \rho _{41}\rho _{23}^{\prime } + \rho_{32}\rho _{14}^{\prime }\right)  & \rho _{33}\rho _{11}^{\prime }+\rho_{44}\rho _{22}^{\prime } & 0 \\
\pm \left( \rho _{41}\rho _{41}^{\prime } + \rho_{32}\rho _{32}^{\prime }\right)  & 0 & 0 & \rho _{33}\rho _{33}^{\prime}+\rho _{44}\rho _{44}^{\prime }
\end{array}%
  \right)  , \label{rhoABphi}
\end{equation}
\begin{equation}
\hat{\rho}_{AB}^{\psi^{\pm }}  \equiv \frac{1}{N_{\psi }} \left(
\begin{array}{cccc}
\rho _{11}\rho _{22}^{\prime }+\rho _{22}\rho _{11}^{\prime } & 0 & 0 & \pm \left(\rho_{14}\rho _{23}^{\prime }+\rho _{23}\rho _{14}^{\prime }\right)  \\
0 & \rho _{11}\rho _{44}^{\prime }+\rho _{22}\rho _{33}^{\prime } & \pm \left(\rho_{14}\rho _{41}^{\prime }+\rho _{23}\rho _{32}^{\prime }\right)  & 0 \\
0 & \pm \left( \rho _{41}\rho _{14}^{\prime } + \rho_{32}\rho _{23}^{\prime }\right)  & \rho _{33}\rho _{22}^{\prime }+\rho_{44}\rho _{11}^{\prime } & 0 \\
\pm \left( \rho _{41}\rho _{32}^{\prime } + \rho_{32}\rho _{41}^{\prime }\right)  & 0 & 0 & \rho _{33}\rho _{44}^{\prime}+\rho _{44}\rho _{33}^{\prime }
\end{array}%
  \right)  , \label{rhoABpsi}
\end{equation}\label{XrhoAB}
\end{subequations}
\end{widetext}
where we have defined the normalization constants,
\begin{eqnarray*}
N_\phi &=&\left( \rho _{11}+\rho _{33}\right) \left( \rho _{11}^{\prime}+\rho _{33}^{\prime }\right) +\left( \rho _{22}+\rho _{44}\right) \left(\rho _{22}^{\prime }+\rho _{44}^{\prime }\right), \\
N_\psi &=&\left( \rho _{11}+\rho _{33}\right) \left( \rho _{22}^{\prime }+\rho_{44}^{\prime }\right) +\left( \rho _{22}+\rho _{44}\right) \left( \rho_{11}^{\prime }+\rho _{33}^{\prime }\right).
\end{eqnarray*}
The probabilities of obtaining each one of the four possible outcomes (\ref{XrhoAB}) are $P_{\phi^\pm}=N_\phi/2$ and $P_{\psi^\pm}=N_\psi/2$.
We can note that the four possible outcomes, $\hat{\rho}^{\phi^\pm}_{AB}$ and $\hat{\rho}^{\psi^\pm}_{AB}$, are $X$-states as well.
This means that the $X$-feature of the input states of pairs $A,C1$ and $B,C2$ is swapped to the state of the pair $A,B$.

The diagonal elements of the outcomes (\ref{XrhoAB}) depend only on the diagonal elements of the input state,
and the off-diagonal elements of (\ref{XrhoAB}) depend on the off-diagonal elements of the input states.
Thus, both terms $\rho_{14}$ and $\rho_{23}$ affect and contribute to the coherence into both subspaces $\mathcal{H}_{00,11}$ and $\mathcal{H}_{01,10}$ of the pair $A,B$.

The two states $\hat{\rho}^{\phi^+}_{AB}$ and $\hat{\rho}^{\phi^-}_{AB}$ are equivalent by means of local unitary operators, say,
$I_A\otimes(|0_B\rangle\langle 0_B|-|1_B\rangle\langle 1_B|)$ or $(|0_A\rangle\langle 0_A|-|1_A\rangle\langle 1_A|)\otimes I_B$;
this means that they have equal amounts of quantum correlation.
Similarly, the states $\hat{\rho}^{\psi^+}_{AB}$ and $\hat{\rho}^{\psi^-}_{AB}$ are equivalent with the same local unitary operators.
However, the two states $\hat{\rho}^{\phi^{\pm}}_{AB}$, in general, are not local-unitarily equivalent to the two $\hat{\rho}^{\psi^{\pm}}_{AB}$ states.
Therefore, in this process the entanglement is distributed probabilistically with concurrence $\mathcal{C}_{AB}^{\phi}$ in the two outcomes $\hat{\rho}^{\phi^{\pm}}_{AB}$ and with $\mathcal{C}_{AB}^{\psi}$ in both states $\hat{\rho}^{\psi^{\pm}}_{AB}$, which, in general, are different.
This asymmetric distribution is reminiscent of the case exposed in Sec. \ref{pure state} for pure states.

From Eqs. (\ref{XrhoAB}) we realize that the matrix elements of $\hat{\rho}_{A,C1}$ are transferred to the outcome states of pair $A,B$ when the state $\hat{\rho}_{B,C2}=|\phi^+_{B,C2}\rangle\langle\phi^+_{B,C2}|$.
In this particular case, the states $\hat{\rho}^{\phi^{\pm}}_{AB}$ are local-unitarily equivalent with the $\hat{\rho}^{\psi^{\pm}}_{AB}$ states respectively by means of the local unitary transformation $I_A\otimes\sigma_x^{B}$ or $\sigma_x^{A}\otimes I_B$.
This means that in this process, all the features of the $X$-state $\hat{\rho}_{A,C1}$ are transferred to the state of the two remote qubits $A$ and $B$.

In order to simplify the analysis and to shed some light on the principal aspects and the scope of this scheme, in the next sections we restring it to the case
with elements of the matrices (\ref{Xstates}) equal, i.e., $\rho^{\prime }_{nm}=\rho_{nm}$.

\section{entanglement swapping} \label{x state entanglement}

In what follows we analyze the entanglement redistribution from $A,C1$ and $B,C2$ pairs to the $A,B$ system, when both initial $X$-states (\ref{Xstates}) are alike, i.e., $\rho^{\prime }_{nm}=\rho_{nm}$.
In this case, the concurrences of the respective outcome states (\ref{rhoABphi}) and (\ref{rhoABpsi}) become
\begin{subequations}\begin{eqnarray}
\!\!\!\!\!\!\!\!\mathcal{C}_{AB}^{\phi}(\!\Delta\!)\!&=&\!\!\frac{2\max\left\{0,g(\!\Delta\!)\!-\!(\rho_{11}\rho_{33}+\rho_{22}\rho_{44})\right\}}
{(\rho_{11}+\rho_{33})^2+(\rho_{22}\!+\!\!\rho_{44})^2\!\!}\!,\\
\!\!\!\!\!\!\!\!\mathcal{C}_{AB}^{\psi}(0)\!&=&\!\!\frac{\max\left\{0,g(0)-2\sqrt{\rho_{11}\rho_{22}\rho_{33}\rho_{44}}\right\}}{(\rho_{11}+\rho_{33})(\rho_{22}+\rho_{44})},
\end{eqnarray}\label{CABs}\end{subequations}
where we have defined the function
\begin{equation}
g(\varphi)=\sqrt{(|\rho_{14}|^{2}+|\rho_{23}|^{2})^{2}-4|\rho_{14}|^{2}|\rho_{23}|^{2}\sin^{2}(\varphi)},
\end{equation}
with $\Delta=\theta_{14}-\theta_{23}$ and $\theta_{nm}$ being the phase of the off-diagonal elements, i.e., $\rho_{nm}=|\rho_{nm}|e^{i\theta_{nm}}$.

The concurrence $\mathcal{C}_{AB}^{\psi}(0)$ does not depend on the phases $\theta_{nm}$ and it is higher than $\mathcal{C}_{AB}^{\phi}(\Delta)$ for all values of $\Delta$.
In addition to having asymmetrically distributed the entanglement over the outcomes, the probability $2P_\psi$ of having $\mathcal{C}_{AB}^{\psi}(0)$ is smaller than the one $2P_\phi$ of having $\mathcal{C}_{AB}^{\phi}(\Delta)$.

The concurrence $\mathcal{C}_{AB}^{\phi}(\Delta)$ reaches its maximal value at $\Delta=0$ and the minimum for $\Delta=\pi/2$.
Taking into account that the relative phases $\theta_{nm}$ can be managed by properly choosing the axes on the Bloch sphere of the qubits,
we can assume $\theta_{14}=\theta_{23}$, which leads to $\Delta=0$.
In this case the concurrences (\ref{CABs}) become
\begin{subequations}
\begin{eqnarray}
\!\!\!\!\!\!\!\!\!\mathcal{C}_{AB}^{\phi}\!\!\!&=&\!\!\!\frac{2\max\left\{0,|\rho_{14}|^{2}\!+\!|\rho_{23}|^{2}\!-\!(\rho_{11}\rho_{33}+\rho_{22}\rho_{44}\!)\!\right\}}
{(\rho_{11}\!+\!\rho_{33})^2+(\rho_{22}\!+\!\rho_{44})^2}\!,\label{Cphi0}\\
\!\!\!\!\!\!\!\!\!\mathcal{C}_{AB}^{\psi}\!\!\!&=&\!\!\!\frac{\max\left\{0,|\rho_{14}|^{2}\!+\!|\rho_{23}|^{2}\!-\!2\sqrt{\rho_{11}\rho_{22}\rho_{33}\rho_{44}}\right\}}
{(\rho_{11}\!+\!\rho_{33})(\rho_{22}\!+\!\rho_{44})}.\label{Cpsi0}
\end{eqnarray}\label{Cpshi0}%
\end{subequations}%
From these expressions we find that the four outcome states (\ref{XrhoAB}) are entangled if the matrix elements of the input states satisfy the inequality
\begin{equation}
|\rho_{14}|^{2}+|\rho_{23}|^{2}>\rho_{11}\rho_{33}+\rho_{22}\rho_{44}, \label{algo}
\end{equation}
whereas the entanglement is present only in two outcome states (\ref{rhoABpsi}) if the following inequalities hold:
\begin{equation}
\rho_{11}\rho_{33}+\rho_{22}\rho_{44}\geq|\rho_{14}|^{2}+|\rho_{23}|^{2}>2\sqrt{\rho_{11}\rho_{22}\rho_{33}\rho_{44}}, \label{ineinf}
\end{equation}
otherwise the four outcomes states are separable.

By making a detailed analysis of the inequalities (\ref{positivity}), (\ref{rhooverrho}), (\ref{eqFORe}), (\ref{ineinf}), and (\ref{algo}) we can asseverate what follows:
\begin{itemize}
  \item[a)] If the input states $\hat{\rho}_{A,C1}$ and $\hat{\rho}_{B,C2}$ are not entangled then $|\rho_{14}|\leq\sqrt{\rho_{22}\rho_{33}}$ and $|\rho_{23}|\leq\sqrt{\rho_{11}\rho_{44}}$.
  The suitably multiplication of the respective terms of these inequalities with those for positivity $|\rho_{14}|\leq\sqrt{\rho_{11}\rho_{44}}$ and $|\rho_{23}|\leq\sqrt{\rho_{22}\rho_{33}}$ leads to $|\rho_{14}|^2\leq\sqrt{\rho_{11}\rho_{22}\rho_{33}\rho_{44}}$ and $|\rho_{23}|^2\leq\sqrt{\rho_{11}\rho_{22}\rho_{33}\rho_{44}}$.
  Adding them we obtain that $|\rho_{14}|^2+|\rho_{23}|^2\leq 2\sqrt{\rho_{11}\rho_{22}\rho_{33}\rho_{44}}$, which means that the right hand side inequality of expression (\ref{ineinf}) is not fulfilled.
  This means that, if the input states lack entanglement, then the four outcome states are separable.

  Additionally, the fact that the concurrence $\mathcal{C}_{in}$ of the input states is different from zero does not guarantee that the outcomes states are entangled.
  Specifically, by considering that $\mathcal{C}_{in}>0$, we find that:
  \item[b)] Only the two outcome states $\hat{\rho}_{AB}^{\psi _{\pm }}$ have entanglement different from zero if
  \[
  \mathcal{C}^{\text{th}}_{\min}<\mathcal{C}_{in}\leq \mathcal{C}^{\text{th}}_{\max}.
  \]
  \item[c)] The entanglement is present in the four outcome states $\hat{\rho}_{AB}^{\psi _{\pm }}$ and $\hat{\rho}_{AB}^{\phi _{\pm }}$ if
  \[
  \mathcal{C}_{in}>\mathcal{C}^{\text{th}}_{\max}.
  \]
  \end{itemize}
The two threshold concurrence values are
\begin{eqnarray}
\mathcal{C}^{\text{th}}_{\min}&=&2(\sqrt{2\sqrt{\rho_{11}\rho_{22}\rho_{33}\rho_{44}}-\min\{|\rho_{14}|^2,|\rho_{23}|^2\}} \nonumber\\
&&-\min\{\sqrt{\rho_{11}\rho_{44}},\sqrt{\rho_{22}\rho_{33}}\}) ,   \\
\mathcal{C}^{\text{th}}_{\max}&=&2(\sqrt{\rho_{11}\rho_{33}+\rho_{22}\rho_{44}-\min\{|\rho_{14}|^2,|\rho_{23}|^2\}} \nonumber\\
&&-\min\{\sqrt{\rho_{11}\rho_{44}},\sqrt{\rho_{22}\rho_{33}}\}).
\end{eqnarray}

These threshold values are decreasing functions of the off-diagonal element $\min\{|\rho_{14}|^{2},|\rho_{23}|^{2}\}$,
which does not affect the amount of $\mathcal{C}_{in}$.
Therefore, we can realize that, when the input states have entanglement, the off-diagonal term $\min\{|\rho_{14}|^{2},|\rho_{23}|^{2}\}$
now plays an important roll for achieving the task of having entanglement in the outcomes.
In fact, it allows to decrease the threshold values and to increase the outcomes entanglement.
In other words, the coherence degree inside the respective subspace helps to increase the entanglement of the outcome states, when the input states have entanglement, of course.

Consequently, the outcome concurrence (\ref{Cpshi0}) reaches the maximum value when there is maximal coherence inside both subspaces $\mathcal{H}_{00,11}$ and $\mathcal{H}_{01,10}$,
namely $|\rho_{14}|=\sqrt{\rho_{11}\rho_{44}}$ and $|\rho_{23}|=\sqrt{\rho_{22}\rho_{33}}$.
Therefore, in this case and for fixed diagonal elements, the highest outcome concurrence is
\begin{equation}
\mathcal{C}_{AB,\max}^{\psi}=\frac{(\sqrt{\rho_{11}\rho_{44}}-\sqrt{\rho_{22}\rho_{33}})^2}
{(\rho_{11}\!+\!\rho_{33})(\rho_{22}\!+\!\rho_{44})},\label{Cmax2}
\end{equation}
and the smallest outcome concurrence becomes
\begin{equation}
\mathcal{C}_{AB,\max}^{\phi}=\max\left\{0,\frac{2(\rho_{11}-\rho_{22})(\rho_{44}-\rho_{33})}{(\rho_{11}+\rho_{33})^2+(\rho_{22}+\rho_{44})^2}\right\}. \label{Cmax3}
\end{equation}
Here, we easily note that $\mathcal{C}_{AB,\max}^{\psi}$ is higher and $\mathcal{C}_{AB,\max}^{\phi}$ is smaller than the input one $\mathcal{C}_{in}$.
Note that (\ref{Cmax3}) is different from zero when $\rho_{11}>\rho_{22}$ and $\rho_{44}>\rho_{33}$ or $\rho_{11}<\rho_{22}$ and $\rho_{44}<\rho_{33}$,
which are conditions for having initial entanglement, in agreement with inequality (\ref{algo}) evaluated for maximal coherence.

It is worth emphasizing that, in order to get in the process maximal outcome entanglement, it is required to manipulate the initial states (\ref{Xstates}) in such a way that $\theta_{14}=\theta_{23}$.
This can be done by applying the local unitary operation $|0\rangle\langle 0|+e^{i(\theta_{14}-\theta_{23})/2}|1\rangle\langle 1|$ to the $C1$ and the $C2$ qubits, before performing the measurement procedure.

Here we have found two threshold concurrence values, conditions for which this scheme allows redistributing entanglement onto two remote qubits.
The entanglement is distributed asymmetrically onto four possible outcomes, thus having probabilities different from zero of increasing and decreasing it with respect to the initial entanglement.
Moreover, the probability of increasing it is smaller than the one for decreasing it.

\section{Examples} \label{rhodifferent}

We show two examples which illustrate the effect produced by the smallest module of the off-diagonal term of the input states on the outcome entanglement.
In the cases with Werner states and with $\alpha$-states the module $\min\{|\rho_{14}|^2,|\rho_{23}|^2\}=0$;
consequently the outcomes demand input states with finite entanglement.
In the case with input $\beta$-states the module $\min\{|\rho_{14}|^2,|\rho_{23}|^2\}\neq0$; accordingly the outcome entanglement arises for all $\beta\neq1/2$ similarly to the input state.

The relation between discord, entanglement of formation (EoF), and linear entropy was addressed in Ref. \cite{AA}.
A. Al-Qasimi and D.F.V. James found the states for which discord takes extreme values for a given entropy or given entanglement.
The upper bound of entanglement-discord relation is given by the $\alpha$-state for $0\leq$EoF$\leq0.620$.
The lower bound of that relation is satisfied by the $\beta$-state for $0\leq$EOF$\leq1$.
Thus, in this section we analyze the outcome states (\ref{Xstates}) when input states are both $\alpha$ or both $\beta$.

Before developing these examples, it is worth taking a gander to the case when the input states are the well-known Werner states \cite{WS}:
$\hat{\rho}(\gamma)=(1-\gamma)I/4+\gamma|\psi^+\rangle\langle\psi^+|$, with $0\leq\gamma\leq 1$.
In this case, the input states are not separable for all $\gamma>1/3$; the two threshold concurrence values are equal,
$\mathcal{C}_{min}^{\text{th}}=\mathcal{C}_{max}^{\text{th}}=\sqrt{(1-\gamma^2)/2}-(1-\gamma)/2$.
Considering the asseveration c) of Sec. \ref{x state entanglement}, the four outcome states are entangled when $\mathcal{C}_{in}(\gamma)=(3\gamma-1)/2>\sqrt{(1-\gamma^2)/2}-(1-\gamma)/2$,
which holds for $\gamma>1/\sqrt{3}$.

\subsection{Starting with $\alpha$-states}

\begin{figure}[t]
\includegraphics[angle=360,width=0.40\textwidth]{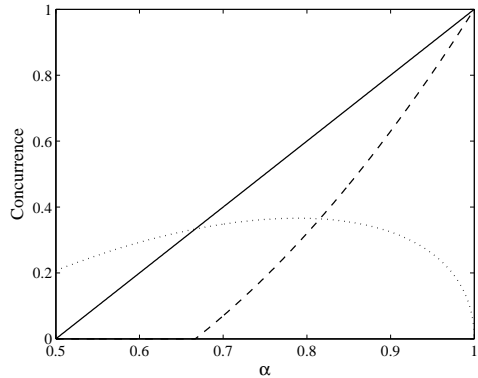}
\caption{Input and outcome concurrences as functions of $\alpha$.
$\mathcal{C}_{in}(\alpha)$ (solid line), $\mathcal{C}_{AB}(\alpha)$ (dashed line), and $\mathcal{C}^{\text{th}}(\alpha)$ (dotted line).}
\label{figure2}
\end{figure}

The $\alpha$ state corresponds to an $X$-type given by
\begin{eqnarray}
\hat{\rho}(\alpha)&=&\frac{1-\alpha}{2}(|\psi^+\rangle\langle\psi^+|+|\psi^-\rangle\langle\psi^-|)\nonumber\\
&& + \alpha|\phi^+\rangle\langle\phi^+|, \label{alpha}
\end{eqnarray}
where the $\alpha$ real parameter goes from $0$ to $1$ and $|\psi^\pm\rangle$ and $|\phi^\pm\rangle$ are the Bell states.
At $\alpha=0$ the $\hat{\rho}(\alpha)$ state is an incoherent classical state and at $\alpha=1$ is the Bell state $|\phi^+\rangle$.
The concurrence $\mathcal{C}_{in}(\alpha)$ of $\hat{\rho}(\alpha)$ is \cite{AA}
\begin{equation}
\mathcal{C}_{in}(\alpha)=\max\{0,2\alpha-1\}.
\end{equation}
The $\hat{\rho}(\alpha)$ state is separable for all $\alpha$ between $0$ and $1/2$ and is entangled for $1/2<\alpha\leq 1$.
In this case, the four outcome states (\ref{Xstates}) are local-unitarily equivalent and given by
\begin{eqnarray*}
\hat{\rho}_{AB}^{\phi^{\pm}}\left(\alpha\right)&=&\alpha\left( 1-\alpha\right)\left(|\psi_{AB}^{+}\rangle\langle\psi_{AB}^{+}|+|\psi_{AB}^{-}\rangle\langle\psi_{AB}^{-}|\right) \\
&&+\left[\frac{\alpha^{2}+\left(1-\alpha\right)^{2}}{2}\pm\frac{\alpha^{2}}{2}\right]|\phi_{AB}^{+}\rangle\langle\phi_{AB}^{+}|  \\
&&+\left[\frac{\alpha^{2}+\left(1-\alpha\right)^{2}}{2}\mp\frac{\alpha^{2}}{2}\right]|\phi_{AB}^{-}\rangle\langle\phi_{AB}^{-}|, \\
\hat{\rho}_{AB}^{\psi^{\pm}}\left(\alpha\right)&=&\sigma_x^{(A)}\hat{\rho}_{AB}^{\phi ^{\pm }}\left( \alpha \right)\sigma_x^{(A)}.
\end{eqnarray*}
The common concurrence of them is
\begin{equation}
\mathcal{C}_{AB}\left( \alpha \right) =\max \{0,\alpha \left( 3\alpha -2\right) \}.
\end{equation}
On the other hand, for the input state (\ref{alpha}) the two threshold concurrence values are equal, $\mathcal{C}_{min}^{\text{th}}=\mathcal{C}_{max}^{\text{th}}=\mathcal{C}^{\text{th}}(\alpha)$, with
\begin{equation*}
\mathcal{C}^{\text{th}}(\alpha)=\sqrt{2\alpha(1-\alpha)}-(1-\alpha),\quad\text{for }\alpha>1/2;
\end{equation*}
therefore, the outcome entanglement arises when $\mathcal{C}_{in}(\alpha)>\mathcal{C}^{\text{th}}(\alpha)$, which holds for $\alpha>2/3$.
Fig. \ref{figure2} illustrates this behavior showing $\mathcal{C}_{in}(\alpha)$, $\mathcal{C}_{AB}(\alpha)$, and $\mathcal{C}_{max}^{\text{th}}$ as functions of $\alpha$.

\subsection{Starting with $\beta$-states}

\begin{figure}[t]
\includegraphics[angle=360,width=0.40\textwidth]{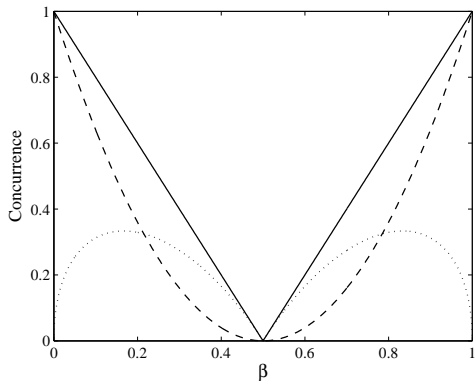}
\caption{Input and outcome concurrences as functions of $\beta$.
$\mathcal{C}_{in}(\beta)$ (solid line), $\mathcal{C}_{AB}(\beta)$ (dashed line), and $\mathcal{C}_{max}^{\text{th}}$ (dotted line).}
\label{figure3}
\end{figure}

The $\beta$-state is a particular $X$-type given by the expression
\begin{equation}
\hat{\rho}(\beta)=\beta|\phi^+\rangle\langle\phi^+|+(1-\beta)|\psi^+\rangle\langle\psi^+|,
\end{equation}
where $\beta$ is a real number going from $0$ to $1$; at both ends it becomes a Bell state and is separable at $\beta=1/2$.
Its concurrence $\mathcal{C}_{in}(\beta)$ is \cite{AA}
\begin{equation}
\mathcal{C}_{in}(\beta)=|1-2\beta|.
\end{equation}
In this case, the outcome states (\ref{XrhoAB}) become
\begin{eqnarray*}
\hat{\rho}_{AB}^{\phi^{\pm}}(\beta)&=&[\beta^2+(1-\beta)^2]|\phi^\pm_{AB}\rangle\langle\phi^\pm_{AB}| \\
&&+2\beta(1-\beta)|\psi_{AB}^\pm\rangle\langle\psi_{AB}^\pm|, \\
\hat{\rho}_{AB}^{\psi^{\pm}}(\beta)&=&\sigma_x^{(A)}\hat{\rho}_{AB}^{\phi^{\pm}}(\beta)\sigma_x^{(A)}.
\end{eqnarray*}
We note that the four outcomes are $\beta$-states with the $\beta$ parameter changed by $\beta^2+(1-\beta)^2$.
The common concurrence of the four outcomes states is
\begin{equation}
\mathcal{C}_{AB}(\beta)=(1-2\beta)^2,
\end{equation}
which is just the square of the input one.
The threshold concurrence values for $\hat{\rho}(\beta)$ are equal, $\mathcal{C}_{\min }^{\text{th}}=\mathcal{C}_{\max }^{\text{th}}=\mathcal{C}^{\text{th}}(\beta)$, and given by
\begin{equation*}
\mathcal{C}^{\text{th}}(\beta)=\left\{
\begin{array}{lll}
\sqrt{\beta(2-3\beta)}-\beta,  &  & \beta <1/2, \\
\sqrt{(3\beta-1)( 1-\beta)}-(1-\beta),  &  & \beta >1/2.
\end{array}%
\right.
\end{equation*}
According to asseveration c) of Sec. \ref{x state entanglement} four outcome states are nonseparable when $\mathcal{C}_{in}(\beta)>\mathcal{C}^{\text{th}}$, i.e., there is outcome entanglement for all $\beta\neq 1/2$.
Fig. \ref{figure3} shows the behavior of $\mathcal{C}_{in}(\beta)$, $\mathcal{C}_{AB}(\beta)$, and $\mathcal{C}_{max}^{\text{th}}$ as functions of $\beta$.

\section{Summary} \label{summary}

In summary we propose a generalization of the simplest scheme for entanglement-swapping beyond pure states.
Basically the generalization is based on performing a von Neumann measurement on two local qubits, but in this case, each one being correlated through an $X$-state with a spatially distant qubit.
This process swaps the $X$ form of the input states without conditions.
When one of the inputs is a Bell state then the other input $X$-state is swapped fully to the remote qubits.
The input entanglements are, in general, partially distributed in the four possible outcome states under certain conditions.
When the input states are equal, we obtain two threshold concurrence values which have to be overcome by the input state entanglement in order for the outcome states to be non-separable.
In addition, in this case, we find that there are two possible amounts of outcome entanglement, one is greater and the other is less than the input entanglement;
the probability of obtaining the greatest outcome entanglement is smaller than the probability of attaining the least one.
Finally, we would like to emphasize that the asymmetric redistribution of the entanglement holds also in the case with pure states,
but the \emph{threshold-concurrence-values} effects are onset only for mixed $X$-states, thus being a consequence of the nonzero decoherence of the input states.

\begin{acknowledgments}
This work was partially supported by grants FONDECyT 1120695 (Chile).
Authors A. M. thanks CONICyT (Chile) and G. G. thanks DAAD RISE Worldwide CL-PH-1999.
\end{acknowledgments}

\end{document}